# Defected photonic crystal-based sensor with enhanced performance

I.M. Efimov, N.A. Vanyushkin, A.H. Gevorgyan

School of Natural Sciences, Far Eastern Federal University, 10 Ajax Bay, Russky Island, 690922 Vladivostok, Russia

**Abstract**

We propose an improved structure of an optical biosensor based on a photonic crystal (PC) with a defect layer (DL), which can detect the concentration of SARS-CoV-2 in water by defect mode (DM) shift. We investigated 4 types of defective PCs with different arrangements of layers inside the perfect PCs and their impact on the performance of the sensor. The sensitivity and amplitude of DM were examined as a function of DL thickness. Also, the peculiarities of edge modes in the presence of DL were investigated. Finally, we obtained a characteristic equation to determine the wavelengths of DMs for an arbitrary 1D PC with an isotropic defect inside.

## 1. Introduction

The number of different types of pathogens increases steadily every year. This poses a threat to humanity, which in turn has led many scientists to look for ways to detect and control pathogens. In the 20th century, it seemed that a universal cure had been found - antibiotics - but viruses and bacteria are adapting. So now we are talking about antibiotic-resistant pathogens, which make treatment and recovery of patients much more difficult. The speed of progression of some lethal viral diseases can be worrying. These challenges dictate the need to develop methods and instruments for the urgent and highly accurate detection of viruses, even at the lowest concentrations. Analytical techniques, such as chemical analysis and also methods using biochemical analyzers, can be used to detect biological agents. In particular, identification procedures based on the real-time reverse transcription-polymerase chain reaction are used for the detection of pathogens such as SARS-CoV-2 [1]. However, this technique requires advanced laboratory testing, expensive equipment, and experienced personnel. Therefore, alternative diagnostic methods are being developed.

An alternative to traditional methods is biosensor methods. Biosensors can be easily miniaturized. They can be used to monitor in real time the concentration of various indicators directly around the patient. The essence of any biosensor consists of two main elements. The first one is a signal transducer, which is used to recognize and translate the signal into a human-readable interface, from the second element. The second element is the biosensor itself. This element provides the necessary parameters for pathogen detection. In particular, optical biosensors have been actively developed recently, because they have high sensitivity to changes in pathogen concentration, can operate in different modes such as transmission, reflection and absorption of light waves, and, unlike biochemical methods, are environmentally friendly, because no chemical reagents that have to be disposed in the process of measurement are used [2-3].

In recent years, biosensors based on photonic crystals (PCs), including those with a defect in the structure, have been of great interest [3-26]. PCs are artificial dielectric structures with periodic refractive index modulation. In recent years, they have attracted much interest due to their unprecedented ability to control the propagation behavior of electromagnetic waves and potential applications in ultra-small all-optical integrated circuits [27]. PCs have the unique property of having a certain range of wavelengths, called the photonic bandgap (PBG), in which an electromagnetic wave cannot propagate through the PC [25]. In practice, this means that if the radiation with a wavelength inside the PBG is incident on the PC, it experiences a strong reflection from the PC. In this way the PC can act as a mirror or an optical filter. If a defect layer (DL) is added to the periodic structure of the PC, the periodicity of the structure is disrupted, causing a change in the transmission and reflection spectra throughout the spectral region. Just as the introduction of DLs into semiconductor superlattices can lead to defect states, DLs in one-dimensional PCs can also create localized defect modes (DMs) within the PBG. The position and shape of the DM depends on the parameters of the DL, such as the thickness and refractive index of the DL. It is this property that underlies our biosensor. In this work, the defect is a layer composed of water with SARS-CoV-2 added. A change in the concentration of the pathogen causes a change in the refractive index of the DL, which in turn shifts the position of the DM, which can be detected by the sensor. The SARS-CoV-2 concentration in the DL is determined by this deviation.

In [28], we investigated an optical PC-based biosensor with a DL. We encountered the following problem: due to the large absorption of the components of the sensor structure in the infrared range, the DM amplitude was too small, which limits the capability of the sensor. In this paper, to solve this problem, the operating range of the sensor was shifted to the visible region where the absorption of water, which constitutes the majority of the DL, is much lower. In addition, we optimized the sensor structure, in particular we moved to a structure with mirror symmetry and investigated the effect of DL thickness on the biosensor sensitivity and DM amplitude in details. Finally, we derived the characteristic equation to determine the DM of an arbitrary PC with a defect and investigated the features of the edge modes in the presence of DL.

## 2. The Theory

The cross-section of the PC we are considering is shown in Fig. 1.

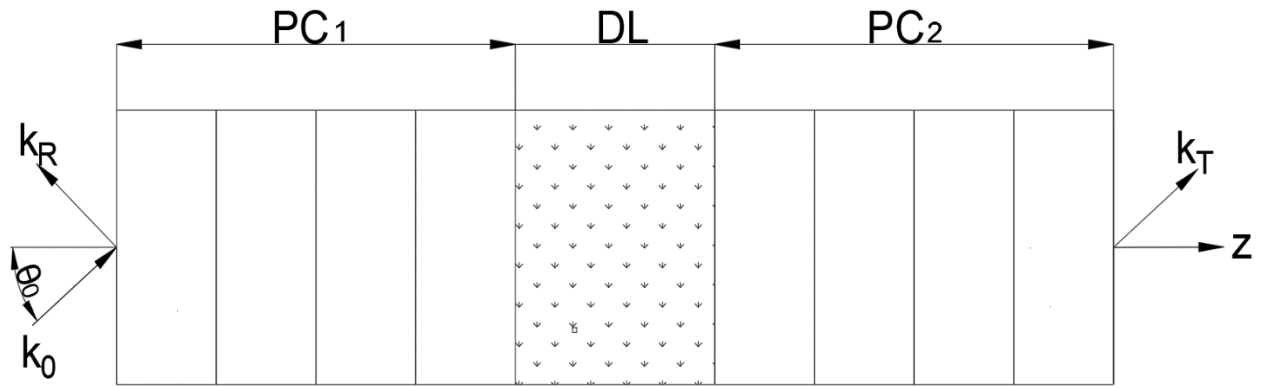

Fig. 1. Schematic diagram of the analyzed structure

$k_0$, $k_R$ and $k_T$ are the wave vectors of the incident, reflected and transmitted waves respectively, $PC_1$, $PC_2$ are the perfect PCs, $\theta_0$ is the angle of incidence, and DL is the defect layer.

Our structure is composed of two periodic perfect PCs with a DL between them. Each perfect PC consists of a sequence of $N$ unit cells each of which is a pair of layers having thickness $h_{1,2}$ and refractive index $n_{1,2}$. The refractive indices were determined by the materials of the layers. For the selection of suitable materials, the following conditions were set: the materials should provide a high difference between the refractive indices of the $n_1$ and $n_2$, layers, be relatively inexpensive and easy to manufacture, and transparent in the spectral range of interest. The thicknesses of the layers were determined from the quarter-wave criterion $Re[n_1(\lambda_b)] \, h_1 = Re[n_2(\lambda_b)] \, h_2 = \lambda_b/4$, where $\lambda_b$ is the wavelength of the center of the PBG. If this condition is met, we obtain the maximum width of the PBG.

The DL itself is a host medium with inclusions. The host medium is water, and the inclusions are SARS-CoV-2 pathogens. The Maxwell-Garnett effective medium approximation [29-30] was used to determine the refractive index of DL through the parameters of the constituent substances. This method is applicable if the macroscopic system is homogeneous and the dimensions of all particles are much smaller than the wavelength [29-30].

Fig. 2 shows different types of structures considered in this work, which include structures with mirror symmetry (a,b,c) and without mirror symmetry (d) in relation to the DL.

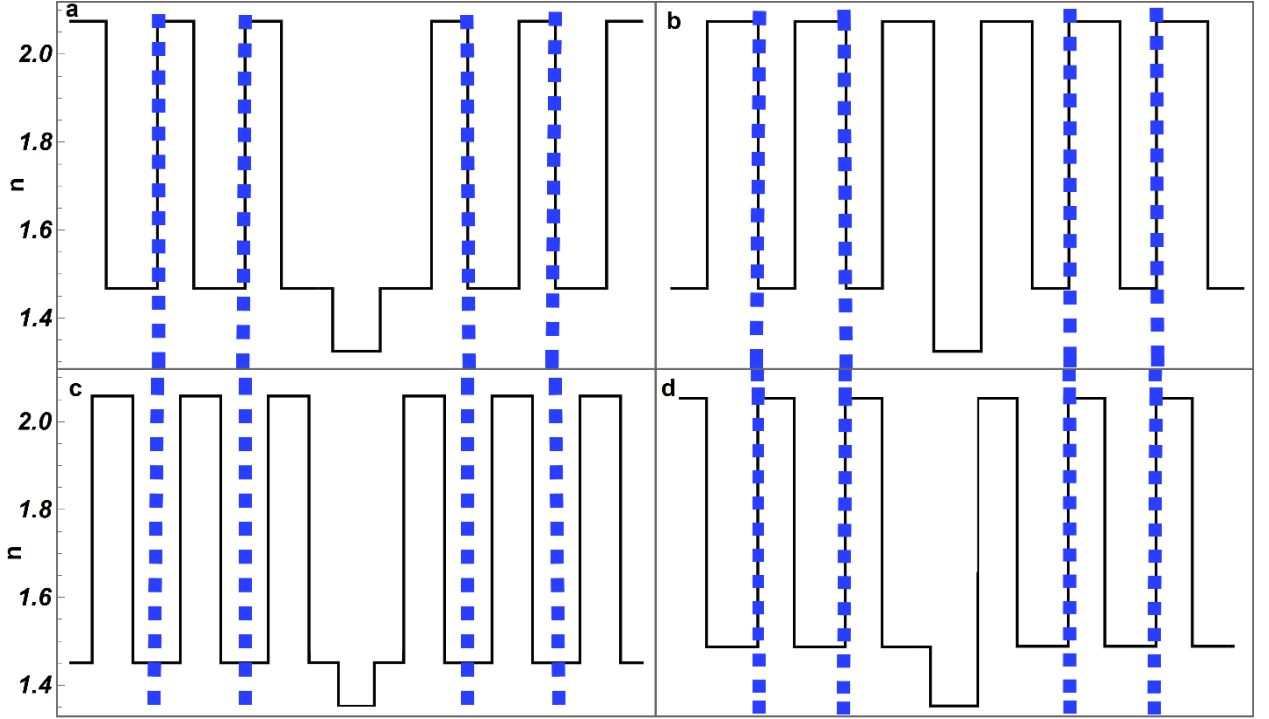

Fig. 2. Types of layer placement within the PC. The blue dashed lines show the unit cell of each perfect PC to the left and right of the DL.

The structure in Fig. 2a (structure 1) is a case of mirror symmetry where the defect is located between two layers with lower refractive indices. The structure in Fig. 2b (structure 2) is a case of mirror symmetry where the defect is between two layers with a higher refractive index. The structure in fig. 2c (structure 3) is a case of both mirror symmetry and symmetric unit cell, where the defect is between two layers with lower refractive indices. The structure in fig. 2d (structure 4) is a case without mirror symmetry which corresponds to the structure in our earlier work [28].

In this paper we used the transfer matrix method [31-33] to calculate the transmission and reflection spectra of the structure under investigation.

**2.1 The characteristic equation for the defected PC eigenmodes.** Finding the DM can be problematic due to the high quality factor (Q-factor) of the mode, which requires calculating the reflection or transmittance spectrum of the structure with sufficiently small wavelength step size. This problem can be solved by using a characteristic equation for the DM.

The transfer matrix of a defective PC can be generally represented as:

$$m = [M_{02} M_2 M_{2d}](M_d)[M_{d1} M_1 M_{10}] = M_{II} M_d M_I, \qquad (1)$$

where $M_1$, $M_d$, $M_2$ are "inner" transfer matrices of PC1, DL and PC2, i.e. these transfer matrices relate the fields at inner sides of their outer boundaries. $M_{02}$ and $M_{10}$ are "boundary" matrices for the interfaces between the perfect PCs and the external environment, i.e. these transfer matrices link relate the fields at different sides of the same boundary. Similarly, the matrices for the boundaries between the perfect PCs and DL are defined as $M_{2d}$ and $M_{d1}$.

In the absence of absorption, any transfer matrix can be represented as [27]:

$$M = \begin{pmatrix} \frac{1}{t^*} & -\frac{r^*}{t^*} \\ -\frac{r}{t} & \frac{1}{t} \end{pmatrix}, \quad (2)$$

where $r$ is reflection coefficient, $t$ is transmission coefficient, the asterisks denote the complex conjugate.

In addition, since the DL is homogeneous, its inner transfer matrix can be further simplified:

$$M_d = \begin{pmatrix} e^{i\varphi} & 0 \\ 0 & e^{-i\varphi} \end{pmatrix}. \quad (3)$$

where $\varphi = \frac{2\pi}{\lambda} n_d h_d \cos(\theta_d)$, and $\theta_d$ is the angle of refraction in the DL.

Now considering (2) and (3) the relation (1) turns into a matrix equation:

$$\begin{pmatrix} \frac{1}{\tau^*} & -\frac{\rho^*}{\tau^*} \\ -\frac{\rho}{\tau} & \frac{1}{\tau} \end{pmatrix} = \begin{pmatrix} \frac{1}{t_{II}^*} & -\frac{r_{II}^*}{t_{II}^*} \\ -\frac{r_{II}}{t_{II}} & \frac{1}{t_{II}} \end{pmatrix} \begin{pmatrix} e^{i\varphi} & 0 \\ 0 & e^{-i\varphi} \end{pmatrix} \begin{pmatrix} \frac{1}{t_I^*} & -\frac{r_I^*}{t_I^*} \\ -\frac{r_I}{t_I} & \frac{1}{t_I} \end{pmatrix}. \quad (4)$$

The coefficients $r$, $t$ are the reflection and transmission coefficients of the respective elements of the structures while $\rho$, $\tau$ are the reflection and transmission coefficients of the entire structure. Since the reflection coefficient $\rho$ is zero for eigenmodes (including DM) [27], then by equating $\rho = 0$ in (4) we obtain the characteristic equation for the DM:

$$e^{2i\varphi} t_I r_{II} + r_I t_I^* = 0 \quad (5)$$

Actually it is more convenient to use this equation in the following form:

$$e^{2i\varphi} r_{II} + r_I \frac{t_I^*}{t_I} = 0 \quad (6)$$

Equations (1-6) are correct if there is no absorption in the structure. When taking into account absorption it is necessary to replace the complex conjugation in the formulas by the inversion of the wave vector $k \rightarrow -k$ [33]. Equation (6) will then take the form:

$$e^{2i\varphi} r_{II(k)} + r_{I(k)} \frac{t_{I(-k)}^*}{t_{I(k)}} = 0 \quad (7)$$

Here the values with indices $k$ and $-k$ are calculated before and after wave vector inversion, respectively. When using the transfer matrix method, the wave vector inversion results in a change in the transfer matrices of all elements of the structure:

$$M_{\pm k} = \begin{pmatrix} \cos(\varphi) & \mp \frac{i}{p} \sin(\varphi) \\ \mp ip \sin(\varphi) & \cos(\varphi) \end{pmatrix}, \quad (8)$$

where $p = n \cdot \cos(\theta)$, $n$ is the refractive index of the layer considered.

In the presence of absorption, equation (6) has no real roots, but the real part of the complex root is very close to the wavelength of the DM. In addition, the imaginary part of the complex root is inversely proportional to the amplitude of the DM.

Thus, when searching for DM, one can look for the zeros of the following function:

$$f(\lambda) = \left| e^{2i\varphi} r_{II(k)} + r_{I(k)} \frac{t_{I(-k)}}{t_{I(k)}} \right| \tag{9}$$

Fig. 3 shows the reflection spectrum of structure 2 (black) and function (8) (blue).

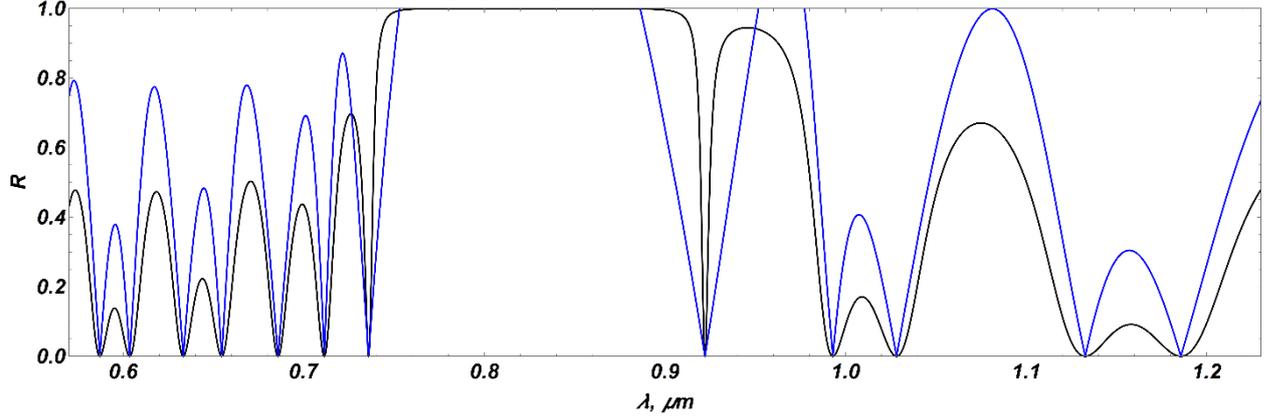

Fig. 3. Spectra of reflection (black) and function (8) (blue) for the structure 2. Parameters of structure $h_1=0.14$ μm, $h_2=0.10$ μm, $h_d=0.44$ μm; $N=10$.

Here one can see that the obtained function has a V-shaped dip which has much broader width than the DM itself. When searching for the DM this dip is easy to detect, so this function will be used later in calculating the sensitivity of the sensor at different DL thicknesses. This method is particularly useful in the case of a very small amplitude of the defect mode, when the reflection minimum is barely different from 1, but the function (8) still reaches zero for the DM.

**2.2 Peculiarities of edge modes in the presence of DL.** In the absence of DL, the edge modes are equidistant in frequency, but the picture changes when DL is added. When considering different types of structures, we found the behavior of the refraction spectrum presented in Fig. 4 for the structure 2. All other types of structures demonstrate a similar dependence.

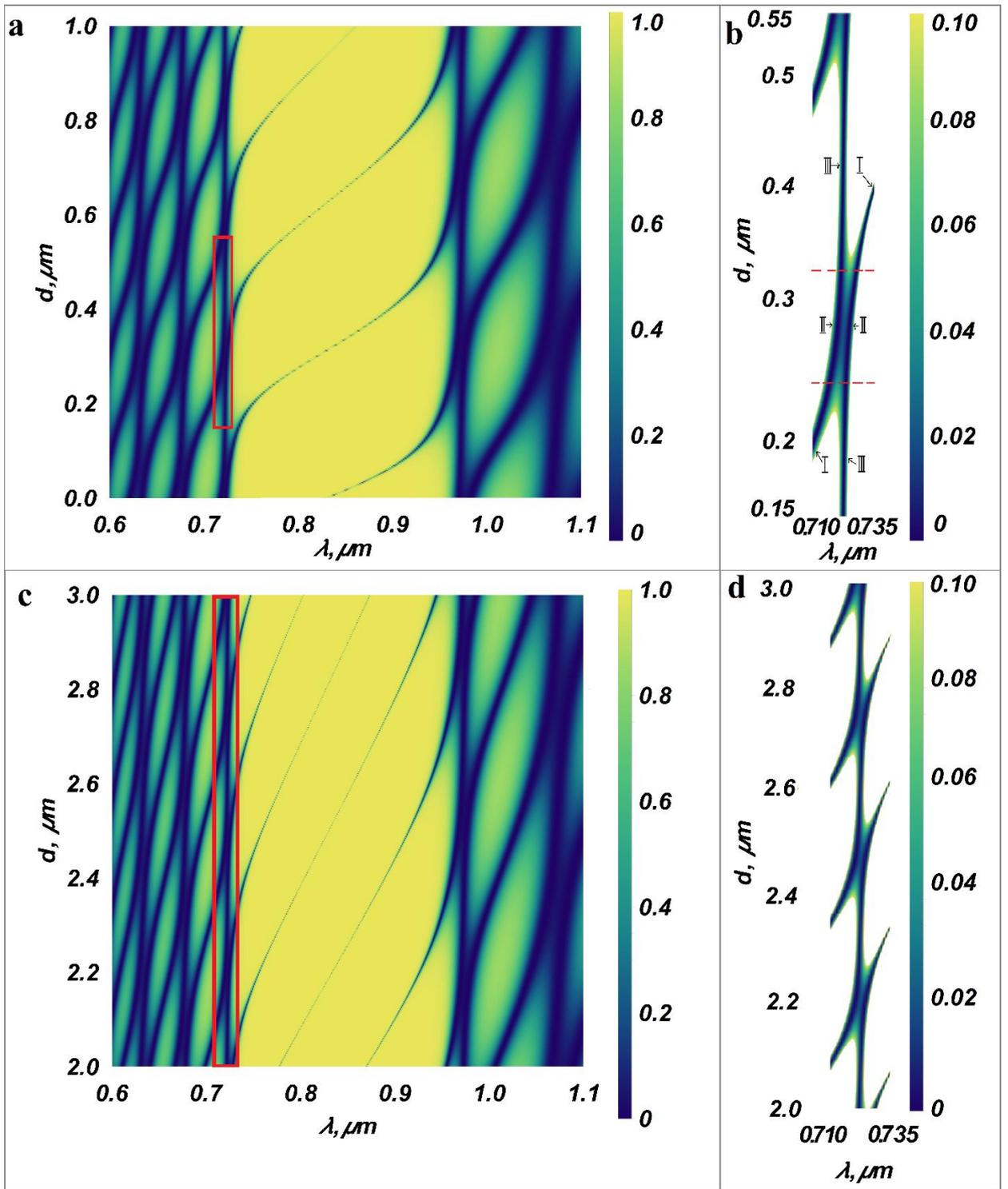

Fig. 4. (a,c) Dependence of reflection spectrum on the DL thickness for the structure 2; (b,d) zoomed in spectrum. All other parameters are as in Fig. 3.

Fig. 4a shows the dependence of reflection spectrum on the thickness of the DL. In the center, the PBG can be seen, bordering a number of edge modes on either side. The defective modes are inside the PBG, and they move as the thickness of the defective layer changes. In Fig. 4a when there is no defect, we still can see a mode inside the PBG, so this mode is "zeroth" mode. As the thickness of the DL increases, the following modes appear, the first, then the second one, and so on.

When considering the Fig. 4a, one can see that when the thickness of the DL changes the edge modes can be divided into 2 types, namely moving and stationary, and the DM refers to the moving ones. As the name implies, moving modes change their position when changing the thickness of

DL, and stationary modes practically do not change their position. Consider now the 2 enlarged modes next to each other in Fig. 4b, for which we can distinguish 3 stages of evolution with increasing thickness of DL. Let us now consider 2 zoomed in modes located next to each other in the Fig. 4b, for which 3 stages can be distinguished. Let us consider the left mode first. At the first stage (I) one can see this edge mode experiences red shift as the DL thickness increases so we call this mode a moving one, but then at the second stage (II) the speed of movement decreases significantly. In addition, it is possible to notice that the distance between the left and right modes is almost constant here. At this stage both modes move in such tied state, until the left mode takes the place of the former stationary mode. After this the third stage (III) starts, where the left mode becomes a stationary one. As a consequence, the mode type change occurs during the second stage. It is important to note that there is no merging of modes at the second stage. If we consider the mode on the right, it has all the same stages, but in the opposite order.

Fig. 4 c,d show the dependence of reflection spectrum on the thickness of the DL at greater values of the DL thickness, when many defect modes arise. As can be seen, the same behavior as above is present here. It is also worth noting that the duration of the stage III (the stationary regime) is constant and does not change at larger thicknesses of DL, but the slope angle of the moving modes in stage I (including DMs) increases, and because of this the number of DMs within the PBG increases too. Let us note, that all the four types of structures, mentioned above, were considered by us, and these effects were found in all of them.

3. Results and discussion

To solve the main problem of the previous work [28], which was the small amplitude of the DM due to the large absorption in the infrared range, it is necessary to minimize the imaginary component of the refractive index in all components of the structure. The main component of our DL is water with a small inclusion of SARS-CoV-2. Thus, to significantly reduce absorption, we moved the working wavelength to the visible range [34], namely around the wavelength 0.8 μm where the refractive index of water is $n_w = 1.328 + i \cdot 1.375 \times 10^{-7}$. Also, this wavelength is situated in the absorption minimum zone of protein compounds (including SARS-CoV-2) in the region of 0.7-1.1 μm [35-38].

The absorption spectrum for SARS-CoV-2 in infrared region was obtained in [39], but now we are working in a shorter wavelength region in which data for SARS-CoV-2 are not available. However, given that there are no protein absorption peaks in this region [35-38], we can use the extrapolation of the refractive index found in [40] from the same spectrum [39].

For this range, a perfect PC was chosen, consisting of layers of SiO2 and TiO2. The refractive indices for these materials are almost constant over the selected range: $n_1 = 2.095 + i \cdot 0$ (TiO2) and $n_2 = 1.469 + i \cdot 0.0013$ (SiO2) [41, 42]. In this paper we used $N = 10$ cells to the left and the same number to the right of the defective layer and we consider the light normal incidence case ($\theta_0 = 0$).

Fig. 5 shows the reflection spectra without DL ($h_d = 0$) for 4 types of structures (in black) and reflection spectra with DL of finite thickness (in blue).

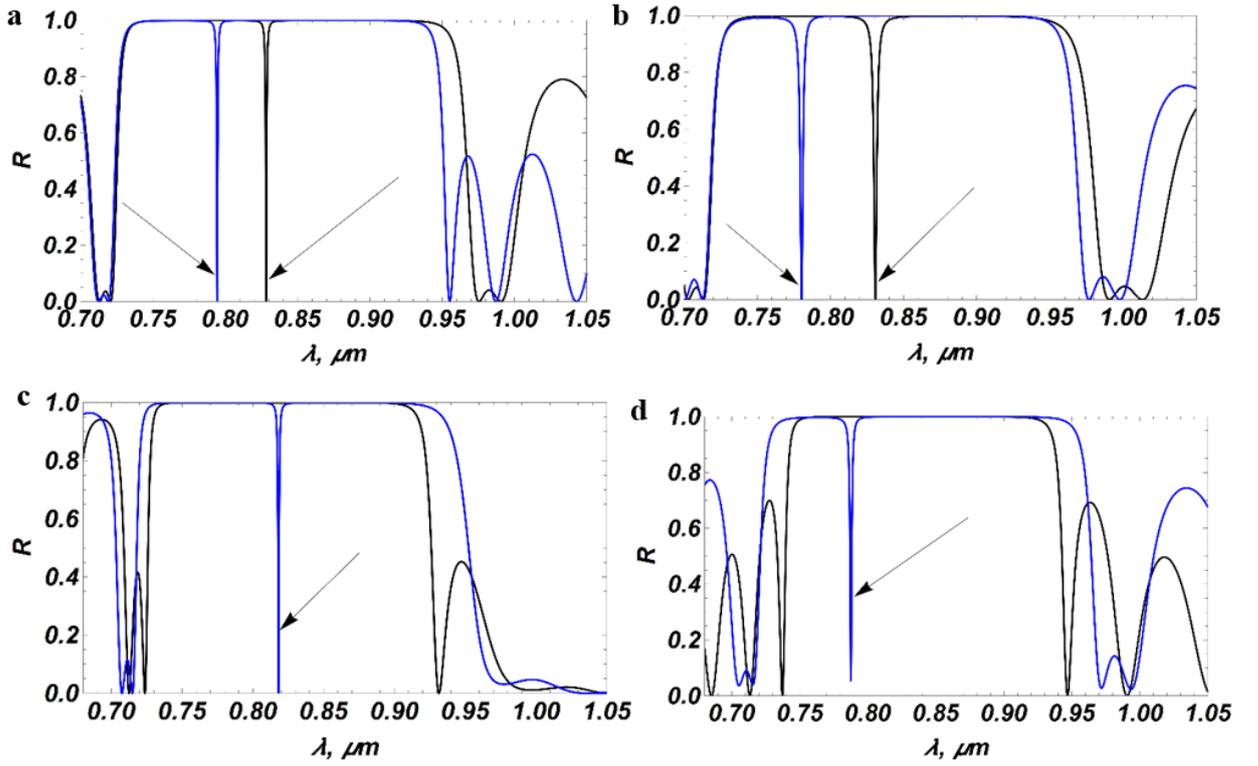

Fig. 5. Reflection spectra of 4 types of structure without a defect layer. (a,b) $h_1=0.14$ µm, $h_2=0.10$ µm, (black) $h_d=0.0$ µm, (blue) $h_d=0.25$ µm; (c) $h_1=0.14$ µm, $h_2=0.5\times2$ µm, (black) $h_d=0.0$ µm, (blue) $h_d=0.10$ µm; (d) $h_1=0.14$ µm, $h_2=10$ µm, (black) $h_d=0.0$ µm, (blue) $h_d=0.10$ µm; $N=10$.

The black arrow indicates DM. Note that in the cases of structures 1 (a) and 2 (b) DM is present despite of the absence of DL. This is the so-called "Zeroth" mode. This is explained by the fact that in the case of mirror symmetry the periodicity of the structure in the center is broken, and this violation plays the role of a defect layer, so the DM appears. It is also worth paying attention to the amplitude of the DM. One can notice that in Fig. 5d, the DM amplitude is smaller than in all the other cases. Therefore, a possible solution to the problem of low DM amplitude may lie not only in a change in the operating spectral range and structure materials, but also in the optimization of the structure geometry.

To compare the structures, it is necessary to consider the sensitivity of the biosensor with respect to changes in the refractive index of DL:

$$S_1 = \frac{d\lambda_d}{dn_d} \qquad (9)$$

Another important parameter is the amplitude of DM, since with too small amplitude there will be no possibility to detect DM.

Fig. 6 shows biosensor sensitivity (9) and DM amplitude as a function of DL thickness (for the first 3 DMs) for the structures under study. Amplitude is represented as T-1, consequently, the T-1 value is proportional to the amplitude of DM.

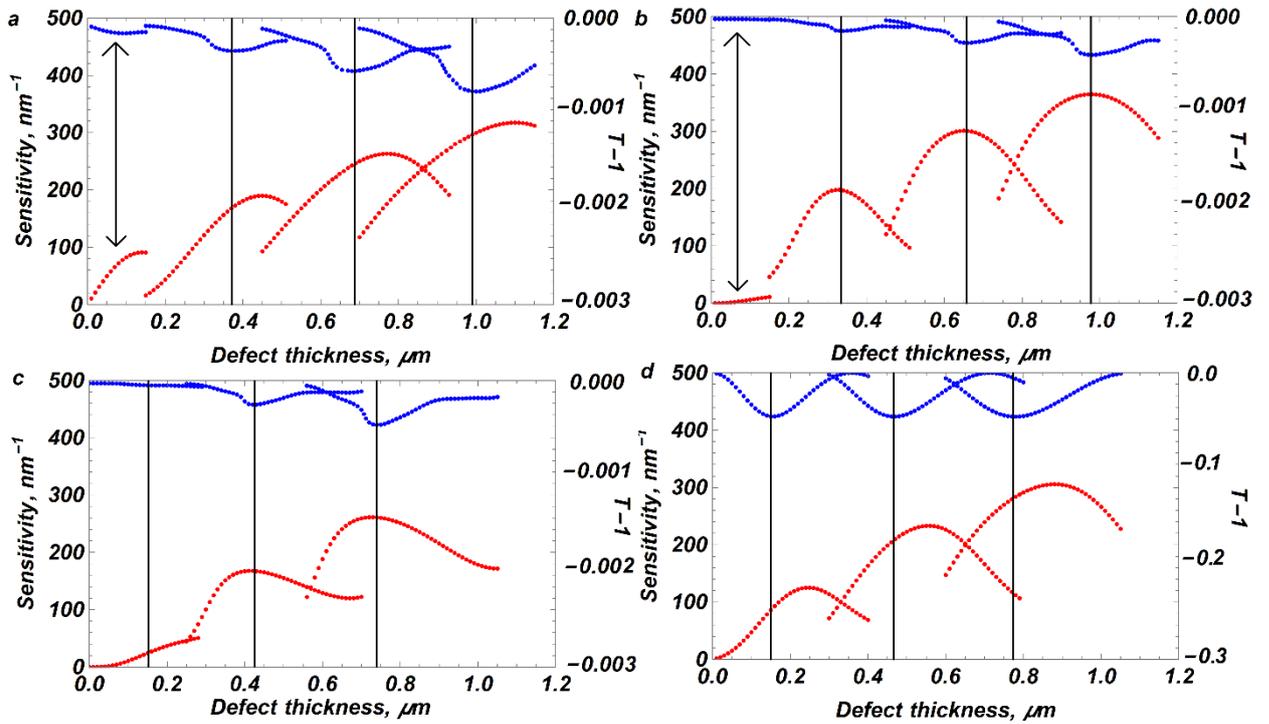

Fig. 6. Sensitivity and amplitude of the biosensor as a function of DL thickness, for 4 structure types: (a) structure 1, (b) structure 2, (c) structure 3, (d) structure 4. Blue points represent DM amplitudes, red points represent sensitivity (9), black lines represent the center of the PBG. The parameters are as in Fig. 2.

It should be noted that in Fig. 6a,b the "zeroth" mode is visible, it is marked with a black arrow. This mode appears due to the fact that the DM in the structures with mirror symmetry is present even in the absence of the defect, as it was mentioned above in this structure the periodicity is broken.

Let us first consider the sensitivity. In all structures we can see the increase of sensitivity with increasing DL thickness, with a general dependence, which is approximately described by the square root [42], but within each DM we can notice that the dependence of sensitivity is not monotonic. These dependencies are similar for all structures. Also, it can be seen that for the structure 2 the maximum of sensitivity is reached in the center of the PBG, while for other structures it does not take place. Overall, the second structure demonstrates the highest sensitivity among all the variants. For this structure, the sensitivity reaches 367 nm$^{-1}$ for the third mod. It is assumed that this is since in the second structure the defect is located between two layers with a higher refractive index, which increases reflection from the borders of DL. As a consequence, the localization within the defect and the sensitivity of the DM increases.

Now consider the amplitude. Structures 1-3 have much higher DM amplitude than structure 4 by 2 orders of magnitude. When considering the general dependence, the DM amplitude decreases with increasing DL thickness, which can be explained by an increase in absorption within the DL. However, for a single DM, we can observe that the changes in the DM amplitude are not monotonic, and, unlike the sensitivity, the minimum is reached in the center of the PBG for all structures. The highest amplitude is also reached at the second structure.

The second structure has the best performance among all 4 structures, so we will investigate it further. Let us now consider the sensitivity of DM to changes in the concentration of CoV-2.

$$S_2 = \frac{d\lambda_d}{d\delta_{CoV}}, \tag{10}$$

where $\delta_{CoV}$ is volume fraction of SARS-CoV-2 in DL.

Fig. 7a shows the DM shift as a function of SARS-CoV-2 concentration, while Fig. 7b shows the wavelength dependence of DM as a function of concentration from 0 to 1% with a linear approximation.

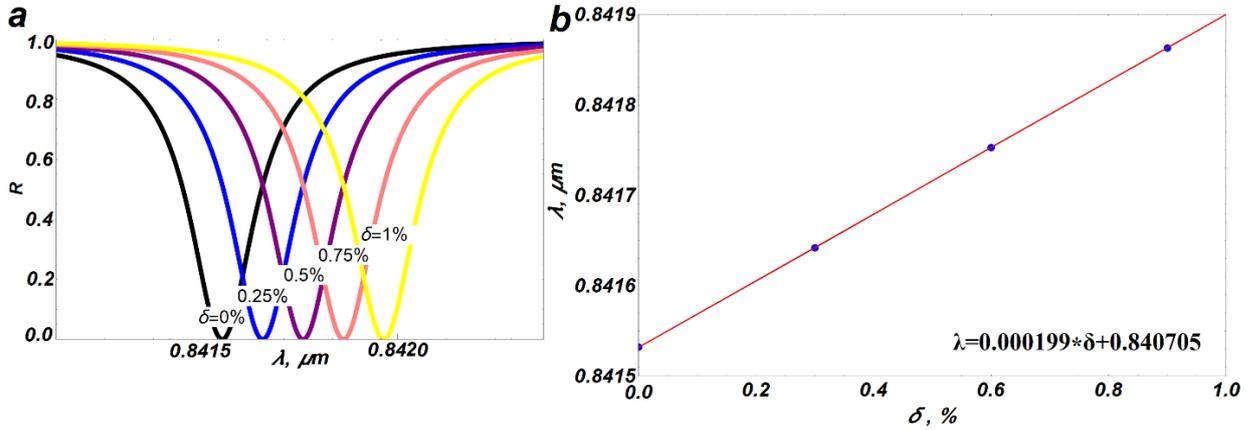

Fig. 7. (a) R spectra with changes in concentration; (b) Dependence of the DM peak on concentration; $h_d$=0.325 μm. All other parameters are as in Fig. 3

Fig. 7a shows the reflection spectra at different concentrations. As the SARS-CoV-2 concentration changes, the DM shifts to the long-wave part of the spectrum, and the Fig. 7b shows this shift. Similar results were obtained in a previous article [28]. A comparative characteristic is presented in Table 1.

Table 1. Comparative characteristic of the sensor sensitivity at different parameters, (A) structure from the article [28], (B) structure 2.

| № | $\lambda_{DM}$ at $\delta_i$ =0, μm, | $S_1, \frac{nm}{RIU}$ | $S_2, \frac{nm}{\%}$ | $\frac{S_1}{\lambda_{DM}}, \frac{1}{RIU}$ | $\frac{S_2}{\lambda_{DM}}, 10^{-3}, \frac{nm}{\%}$ | Q-factor, $\delta = 0$ |
|---|---|---|---|---|---|---|
| A | 5.2930 | 1020 | 1.14 | 0.193 | 0.210 | 300 |
| B | 0.84153 | 208 | 0.21 | 0.247 | 0.250 | 627 |

It can be seen that the sensitivities  and  are higher in structure (A), but this is due to the fact that in the previous work the DM was in the long-wavelength part of the spectrum. When considering the sensitivity normalized to the wavelength of the DM, it turns out that the new structure (B) is much more efficient. It is also worth noting a significant increase in the Q-factor, which favorably affects the performance of the sensor, since it is easier to detect the shift of the DM.

**Conclusion**

In conclusion, we considered an optical biosensor capable of determining the concentration of the pathogen SARS-CoV-2 in water. Four basic types of geometric arrangement of the layers inside a perfect PC were considered. The variants with mirror symmetry and without mirror symmetry were considered. It was found that the variant of the structure without mirror symmetry has the lowest DM amplitude.

The highest sensitivity values are obtained for the structure with mirror symmetry where the DL is located between two layers with a high refractive index (structure 2). For this structure, the sensitivity reaches 367 nm$^{-1}$ at a DL thickness of 0.96 μm.

It was shown that, in general, the sensitivity of the sensor increases with increasing thickness of the defect layer approximately as a square root (see, also [43]), but the dependence for each individual DM is nonmonotonic and reaches its maximum in the center of the PBG. The general pattern for the amplitude is the opposite, with the DM amplitude decreasing as the thickness of the DS increases. However, when considering a particular DM, the amplitude reaches its minimum in the center of the PBG and increases as it moves toward its edges.

A characteristic equation for finding the DM wavelength of an arbitrary 1D PC with a homogeneous defect inside was derived. This equation is particularly useful in finding the DM in the case of very small DM amplitude when the reflection minimum is weakly different from 1, and defect mode halfwidth is very small. Let us note, that this equation describes the edge modes behavior too.

The peculiarities of the edge modes in the presence of DM were investigated. It was shown that with increasing of DL thickness the edge modes and DM are moving, but there are edge modes, which practically do not change their position. The 3 stages were distinguished for each mode, in particular each mode passes a stage of a stationary mode, a coupled mode, and a moving mode. It is important to note, that in the second stage the modes do not merge and move together, as one whole.

The results of this work allow to improve the efficiency of optical sensors based on PC with DL, including biosensors (including gas sensors, or liquid sensors) and other detection devices.

**Acknowledgements**
The work was supported by the Foundation for the Advancement of Theoretical Physics and Mathematics "BASIS" (Grant № 21-1-1-6-1).

**Conflict of interest**
No potential conflict of interest was reported by the authors.